\documentclass{jac}

\usepackage{graphicx}
\usepackage[usenames]{color}

\theoremstyle{plain}

\usepackage[latin1]{inputenc}
\usepackage{stmaryrd}
\usepackage[fixamsmath, disallowspaces]{mathtools}
\newcommand{\Nset}{\mathbb{N}}
\newcommand{\Zset}{\mathbb{Z}}
\newcommand{\Qset}{\mathbb{Q}}
\newcommand{\CA}{\mathcal{C}}
\newcommand{\DFA}{\mathcal{F}}

\newcommand{\LL}{\mathcal{L}}

\newcommand{\W}{\mathcal{W}}
\newcommand{\Cc}{\mathfrak{C}}
\newcommand{\inter}[2]{\llbracket #1, #2\rrbracket}
\newcommand{\Id}{\mathrm{Id}}
\newcommand{\Qa}{Q_\mathrm{a}}
\newcommand{\QQa}{Q'_\mathrm{a}}

\begin{document}

\title[From 1DIDFA to Real-time CA]{A Simulation of Oblivious Multi-head One-way Finite Automata by Real-time Cellular Automata}

\author
{A. Borello}{Alex Borello}
\address
{Laboratoire d'Informatique Fondamentale de Marseille\\
	39 rue Frédéric Joliot-Curie\\ 13453 Marseille, France}  
\email{alex.borello@lif.univ-mrs.fr}  




\begin{abstract}
  \noindent In this paper, we present the simulation of a simple, yet significantly powerful, sequential model by cellular automata. The simulated model is called oblivious multi-head one-way finite automata and is characterised by having its heads moving only forward, on a trajectory that only depends on the length of the input. While the original finite automaton works in linear time, its corresponding cellular automaton performs the same task in real time, that is, exactly the length of the input. Although not truly a speed-up, the simulation may be interesting and reminds us of the open question about the equivalence of linear and real times on cellular automata.
\end{abstract}

\maketitle

\section{Introduction}\label{s:intro}

Cellular automata (CA for short), first introduced by J.~von~Neumann~\cite{VonNeumann:1966} as self-replicating systems, are recognised as a major model of massively parallel computation since A.~R.~Smith, in 1969, used this Turing-complete model to compute functions \cite{Smith:1971:JACM}. Their simple and homogeneous description as well as their ability to distribute and synchronise the information in a very efficient way contribute to their success. However, to determine to what extent CA can fasten sequential computation is not a simple task.

As regards specific sequential problems, the gain in speed by the use of CA is manifest~\cite{Atrubin:1965:EC, Cole:1969:TC, Culik:1989:IPL}. But when we try to get general simulations, we have to face the delicate question of whether parallel algorithms are always faster than sequential ones. An inherent difficulty arises from the fact that efficient parallel algorithms make often use of techniques that are radically different from the sequential ones. There might also exist a faster CA for each singular sequential solution whereas no general simulation exists.

Hence, no surprise: the known simulations of Turing machines by CA provide no parallel speed-up. The early construction of Smith~\cite{Smith:1971:JACM} simulates one step of the Turing machine by one step of the CA. Furthermore, no faster simulations have been reported yet, even for almost all restricted variants. In particular, we do not know whether any finite automata with $k$~heads can be simulated on CA in less than $O(n^k)$ steps, which is the sequential time complexity.

We will not give answers to such issues here, but we shall examine in this context a simple sequential model, called oblivious multi-head finite automata. This device was introduced by M.~Holzer in~\cite{Holzer:2002:TCS} as multi-head finite automata with an additional constraint of obliviousness: the trajectory of the heads only depends on the length of the input. As emphasised in~\cite{Holzer:2002:TCS}, such finite automata lead to significant computational power: they characterise parallel complexity NC$^1$. Their properties have been further discussed in~\cite{HKM:2008:CSP}.

We will focus on the one-way version of this model, that is, for which the reading heads can only move forward (that makes it strictly less powerful). While no true speed-up can be hoped for, as these one-way finite automata already perform their task in linear time, we will describe a simulation of them by real-time CA, that is, CA working in linear time with a multiplicative constant equal to~$1$. Whereas specifying this constant is usually irrelevant, CA represent a particular case amongst models of computation, as we do not know whether linear and real times are equivalent for it.

The article is organised as follows: section~\ref{s:def} introduces the two models considered, section~\ref{s:prel} displays some of their features and abilities and section~\ref{s:sim} presents the simulation algorithm.

\section{Definitions}\label{s:def}

\subsection{Multi-head finite automata}\label{ss:d:dfa}

Given an integer $k\geq 1$, a one-way $k$-head finite automaton is a finite automaton reading an input word using $k$~heads that can move to the right or stand still.
\begin{definition}
	A (deterministic) \emph{one-way multi-head finite automaton}  ($1$DFA($k$) for short) is a septuple $(\Sigma$, $Q, \lhd, q_0, \Qa, k, \delta)$, where $\Sigma$ is a finite set of \emph{input symbols} (or \emph{letters}), $Q$ is a finite set of \emph{states}, $\lhd\notin\Sigma$ is the (right) \emph{end-marker}, $q_0\in Q$ is the \emph{initial} state, $\Qa\subseteq Q$ is the set of the \emph{accepting} states, $k\geq 1$ is the \emph{number of heads} and $\delta : Q\times (\Sigma\cup\{\lhd\})^k\to Q\times\{0, 1\}^k$ the \emph{transition function}; $1$ means to move the head one letter to the right and $0$ to keep it on its current letter. For the heads to be unable to move beyond the end-marker, we require that if $\delta(q, a_1, \dots, a_k) = (q', m_1, \dots, m_k)$, then for any $i\in \inter{1}{k}$, $a_i = \lhd\Rightarrow m_i = 0$.
\end{definition}
A \emph{configuration} of a $1$DFA($k$) on an input word $w\in\Sigma^n$ at a certain time~$t\geq 0$ is a couple $(p, q)$ where $p\in\inter{0}{n}^k$ is the position of the multi-head and $q$ the current state. The computation of such a device on this input word starts with all heads on the first letter, and ends when all heads have reached the end-marker. If the current state is then within $Q_a$, the word is said to be accepted, otherwise it is rejected. The language $L(\DFA)$ \emph{recognised} by a $1$DFA($k$) $\DFA$ is the set of the words accepted by $\DFA$. One can notice a $1$DFA($k$) ends its computation in linear time.

We will focus now on data-independent $1$DFA ($1$DIDFA), a particular class of $1$DFA for which the path followed by the heads only depends on the length of the input word, not on the letters thereof.
\begin{definition}
	Given $k\geq 1$, a $1$DFA($k$) $\DFA$ is said to be \emph{oblivious} (or \emph{data-independent}) if there exists a function $f_\DFA : \Nset^2\to\Nset^k$ such that the position of its multi-head at time $t\in\Nset$ on any input word $w$ is $f_\DFA(|w|, t)$.
\end{definition}

\subsection{Cellular automata}\label{ss:d:ac}

A cellular automaton is a parallel synchronous computing model consisting of an infinite number of finite automata called \emph{cells} which are distributed on $\Zset$ and share the same transition function, depending on the considered cell's previous state as well as its two neighbours'.
\begin{definition}
	 A \emph{cellular automaton}  is a quintuple $(\Sigma, Q, \#, \Qa, \delta)$, where $\Sigma$ is the finite set of \emph{input symbols} (or \emph{letters}), $Q\supset\Sigma$ is the finite set of \emph{states} and $\delta : Q^3\to Q$ the \emph{transition function}\footnote{Notice CA are defined herein with the standard neighbourhood of radius~$1$, that is, such that the state of a cell at time~$t + 1$ depends on the states at time~$t$ of this same cell and its two nearest neighbours.}. $\#\in Q\setminus\Sigma$ is a particular \emph{quiescent} state, verifying $\delta(\#, \#, \#) = \#$. $\Qa\subseteq Q$ is the set of the \emph{accepting} states.
\end{definition}
A \emph{configuration} is a function $\Cc : \Zset\to Q$. A \emph{site} is a cell at a certain time step of the computation we consider; $\langle c, t\rangle$ will denote the state of the site $(c, t)\in\Zset\times\Nset$. The computation of a CA~$\CA$ on an input word $w$ of size~$n\geq 1$ starts at time~$0$ with all cells in state~$\#$ except cells $0$ to $n - 1$ where the letters of the word are written. This is the initial configuration $\Cc_w$ associated to $w$. Then the cells update in parallel their respective states according to $\delta$: for all $(c, t)\in\Zset\times\Nset$, $\langle c, t + 1\rangle = \delta(\langle c - 1, t\rangle, \langle c, t\rangle, \langle c + 1, t\rangle)$.

This input word is accepted in time $t\geq n$ if and only if cell~$0$ (the origin) is in an accepting state at time~$t$. The language $L_\tau(\CA)$ \emph{recognised} by the automaton in time $\tau : \Nset\to\Nset$ is the set of the words~$w$ it accepts in time $\tau(|w|)$. If $\tau$ is the identity function $\Id$, $L_\tau(\CA)$ is said to be recognised in \emph{real time}.

Real time represents for CA the most simple time complexity that is nontrivial, in the sense it is the minimal time required for the output to depend on all letters of the input. Yet, it is significantly powerful, as we do not even know whether linear time can achieve strictly more. Real time had already been evoked in \cite{Smith:1971:JACM}.

\section{Preliminaries}\label{s:prel}

We would like to simulate a $1$DIDFA on a CA as fast as possible. A computation of a general $1$DFA requires a number of time steps that is linear in the size of the input word. Whereas it is rather easy for a CA to simulate such a device in linear time, there is a priori no obvious way to reduce this time bound. But we can do it in the case of DIDFA by taking the constraint of obliviousness into account. Though, before performing such a simulation, we should detail some useful features of DIDFA and CA.

\subsection{Some features of multi-head finite automata}\label{ss:p:dfa}

Let $\DFA = (\Sigma, Q, \lhd, q_0, \Qa, k, \delta)$ be a $1$DIDFA, $n\geq 1$ be an integer and $w\in\Sigma^n$ be a word of size~$n$. Let us look at the computation of $\DFA$ on input word~$w$. For the multi-head is composed of $k$~heads, it can be regarded as a device moving one point at a time in any direction within the set $\W = \inter{0}{n}^k$.

As $\DFA$ is data-independent, we can separate the path $P$ taken by the multi-head from the consecutive states of the automaton (depending on the letters of $w$). In other words, we can take a look at the path of the multi-head on input word $a^n$, for any $a\in\Sigma$; it will be the same for $w$. Hence, the trajectory will become periodic after at most $|Q|$ moves, until one head reaches an end-marker. Then, while the latter head does not move any longer, after another $|Q|$ moves the trajectory will become periodic again, and so on until all heads have reached the end of the input word. The key points of $\W$ where a head reaches the end-marker will be useful to us and denoted as finite sequence $\cramped{(p_i)_{i\in\inter{0}{k}}}$, with $p_0 = (0, \dots, 0)$ and $p_k = (n, \dots, n)$.

\subsubsection*{Some notations}

For convenience, we number the heads such that for all $i\in\inter{0}{k - 1}$, head~$i$ is the one that reaches the end-marker as the multi-head arrives at key point~$p_{i + 1}$. For all $i\in\inter{0}{k}$ and all $j\in\inter{0}{k - 1}$, we denote the $(j + 1)$-th coordinate of $p_i$ by $p_{i, j}$, and if $i < k$ name $P_i\subseteq P$ the portion of trajectory that lies between $p_i$ and $p_{i + 1}$.
\begin{figure}[!ht]
	\centering
	\includegraphics{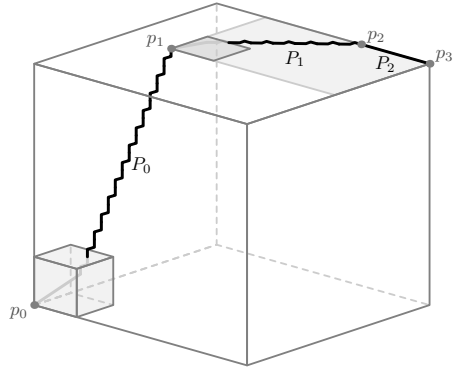}
	\caption{A representation of $\W$ for $k = 3$. The periodic parts of the path of the multi-head are drawn in black.}
	\label{path}
\end{figure}

\subsection{A few basic techniques on cellular automata}\label{ss:p:ca}

A given computation of a CA can be easily represented by drawing successive configurations each one above its predecessor. We thus obtain a \emph{space-time diagram}, composed of sites, of which we only need to represent those in a non-quiescent state.

We will often have to perform several rather independent computations at the same time; this can easily be done by a `product' automaton which works with a finite number of \emph{layers}, each one of which supports a specific computation. Although rather independent, the layers can communicate between one another to exchange information, as any cell can see all of them.

\subsubsection*{Compression of the input word}

\begin{figure}[!ht]
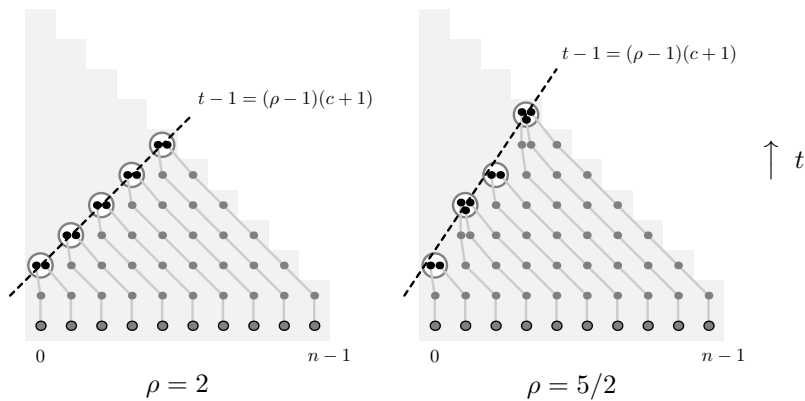

	\centering
	\begin{tabular}{cc}
		\includegraphics{Pix/pix.3} & \includegraphics{Pix/pix.2}\\
	\end{tabular}
	\caption{Schematic space-time diagrams during which input word~$w$ is compressed by rational factor~$\rho$. Each sequence of linked dots represent a letter of $w$. The sites containing the compressed version of $w$ are encircled. Notice that even though it seems the letters could be shifted one time step earlier, this first step is in fact used to mark the last letter; it is necessary because of rounding issues.}
	\label{comp}
\end{figure}

In section~\ref{s:sim}, we will need to compress the input by some rational factor $\rho\geq 2$. This is easy to do with a CA. It consists in having the input word written on the (discrete) straight line of equation $t - 1 = (\rho - 1)(c + 1)$, where $t$ represents the time and $c$ a cell, as shown on fig.~\ref{comp}. As the concerned sites `know' that they lie on this straight line, a computation using the compressed input word can then occur within the triangle of real time (in light grey on fig.~\ref{comp}).

\subsubsection*{Acceleration by a constant}

For any constant $T\in\Nset$ and any CA $\CA$, there exists a CA $\CA'$ such that $L_\Id(\CA') = L_{\Id + T}(\CA)$. In other words, to prove that a given language is CA-recognisable in real time, it suffices to exhibit a CA recognising it in time $\Id + T$. For more details, one can refer to \cite{MazoyerReimen:1992:TCS}.

\section{Simulation}\label{s:sim}

\begin{theorem}\label{THM}
 Given $k\geq 1$, for any $1$DIDFA($k$) $\DFA$ recognising a language $\LL$, there exists a CA $\CA$ recognising $\LL$ in real time.
\end{theorem}
The rest of this paper will be devoted to the proof of this theorem. We assume now that we have a $1$DIDFA($k$) $\DFA = (\Sigma$, $Q, \lhd, q_0, \Qa, k, \delta)$. We will define a CA $\CA = (\Sigma, Q', \#, \QQa, \delta')$ such that $L_\Id(\CA) = \LL$. Instead of giving the full description of its state set and transition function, we will describe its behaviour on an arbitrary input word $w\in\Sigma^n$, given an integer $n\geq 1$. Within this coming description (and similarly in the whole article) the terms `constant' and `finite' refer to quantities that do not depend on $n$.

\subsection{Principle}\label{ss:s:princ}

The general principle of the simulation is rather simple: instead of having $k$~heads moving along $w$, we will have (at least) $k$ copies of $w$ shifted over a segment~$S$ of sites (of strictly increasing time steps) so that each site sees the correct letters of $w$. Moreover, the letters for each head will be seen in reverse order compared to what $\DFA$ does.

\begin{figure}[!ht]
	\centering
	\includegraphics{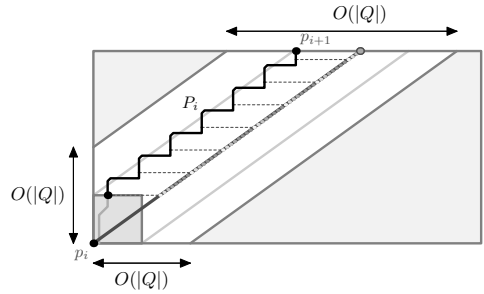}
	\caption{$P_i$ lies within a band of width~$O(|Q|)$, here drawn in white. It can hence be assimilated to a (discrete) straight line, provided a counter (within the shifting copies of $w$ during the execution of $\CA$) indicates for each point of this line the corresponding position within the period of $P_i$. Notice that although the band can broaden as $i$ increases, this index only rises up to a constant value, so that the maximal width $K$ remains bounded independently of the size of the input.}
	\label{band}
\end{figure}

Each part $P_i$ of the trajectory of the multi-head can be assimilated to a discrete straight line, with no aperiodic part. Indeed, as illustrated in fig.~\ref{band}, the distance (in letters) between any point of $P_i$ and the point of this line corresponding to same time step is bounded by some value $K = O(|Q|)$. Thus, during the execution of $\CA$ over $w$, before doing anything, all cells bearing the input will gather the letters of their $K$ nearest neighbours. This is done in time~$K$.

\subsection{Key sites}\label{ss:s:key}

We will set $S = \{(c, n - 1 - c + T)~:~c\in\inter{0}{n - 1}\}$, where $T$, which is to be defined (cf. subsection~\ref{ss:s:comp}), is an integer greater than $K$ that does not depend on $n$. The result of the execution is to appear on site $s_0 = (0, n - 1 + T)$. To know which speed the copies of $w$ should be shifted at over each site of $S$, the latter segment should be divided into parts $S_i$, each one of which corresponds to part $P_i$ of $P$. In other words, we want to mark some key sites $s_i = (c_i, n - 1 -c_i + T)\in S$ that represent key points $p_i\in P$. The main difficulty is that key cell~$c_i$ has to represent coordinate $p_{i, j}$ for any head~$j$.

For this purpose, we observe first that for all $(i, j)\in\inter{0}{k}\times\inter{0}{k - 1}$, since each part of $P$ is as illustrated in fig.~\ref{band}, there exists $\alpha_{i, j}\in\Qset\cap [0, 1]$ such that $|p_{i, j} - \alpha_{i, j}n|\leq K$, whatever the size $n$ of the input. One can notice that we automatically have $\alpha_{0, j} = 0$ and $\alpha_{i, j} = 1$ for all $i > j$, and that $(\alpha_{i, j})_i$ is an increasing sequence for all $j$.

Then, we provisionally assume that $\alpha_{j, j} = 0 \Rightarrow j = 0$, and set key cell $c_i = \lfloor\alpha_i n\rfloor$, where $\alpha_i = \frac 1 2\prod_{j = i}^{k - 1}\alpha_{j, j}$. The case wherein there exists some $j$ that does not verify this hypothesis will be treated in subsection~\ref{ss:s:adjust}.

\begin{figure}[!ht]
	\centering
	\includegraphics{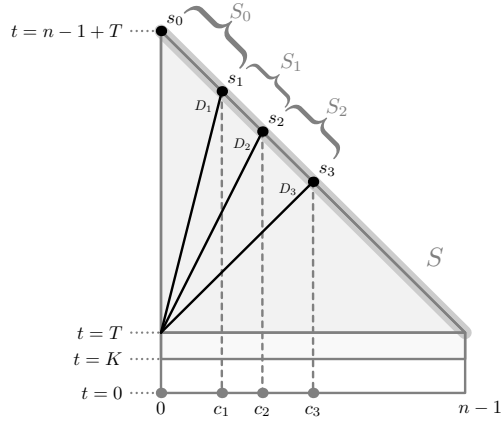}
	\caption{Schematic space-time diagram of the marking of cells $c_i$, for $k = 3$. Notice that $(c_i)_i = (\lfloor\alpha_i n\rfloor)_i$ is always an increasing sequence (since $\alpha_i = \alpha_{i, i}\alpha_{i + 1}\leq\alpha_{i + 1}$), with $c_0 = 0$ and $c_k = \lfloor\frac n 2\rfloor$.	}
	\label{pii}
\end{figure}

Now, how to mark site~$s_i$? No trouble if $i = 0$, as $c_0$ is the origin. If $i > 0$, it is also feasible: it suffices to send a signal from the origin at speed~$\varsigma_i = \cramped{\frac{\alpha_i}{1 - \alpha_i}}\leq 1$ (cf.~fig.~\ref{pii}). Note that in the definition of $\alpha_i$, we have divided by $2$ in case some key cells would be too far from the origin to be marked in time (in CA configurations, information cannot travel at speed of absolute value strictly greater than $1$). All our computation has hence to be performed within half as much space than what $S$ provides. In any case, the definition of $\alpha_i$ is based on the assumption that the copies of the input shifting over $S_i$ are compressed versions of $w$.

\subsection{Compression of the input}\label{ss:s:comp}

For each $i\in\inter 1k$, we want to compress input word $w$ (on a specific layer~$\ell_{i - 1}$ corresponding to head~$i - 1$) by factor $\cramped{\frac 1{\alpha_i}}$ as illustrated in fig.~\ref{comp}, that is, on some straight line $D_i$ of direction vector $(1, \cramped{\frac 1{\varsigma_i}}) = (1, \cramped{\frac1{\alpha_i}} - 1)$. One can notice we are able to choose $D_i$ such that it crosses the origin at any time $t > \lfloor\cramped{\frac 1{\varsigma_i}}\rfloor$. Thus, we will make all such lines cross the origin at the same time $T\in\Nset$. As $(\cramped{\frac 1{\varsigma_i}})_i$ is a decreasing sequence and as we have done some computations in time $K$ beforehand, we set $T = K + \lfloor\cramped{\frac 1{\varsigma_1}}\rfloor$. Hence, we have finally set $D_i$ to be the line of equation $t - T = \cramped{\frac c{\varsigma_i}}$ (cf. fig.~\ref{pii}).

\subsection{Shift of the input}\label{ss:s:shift}

\begin{figure}[!ht]
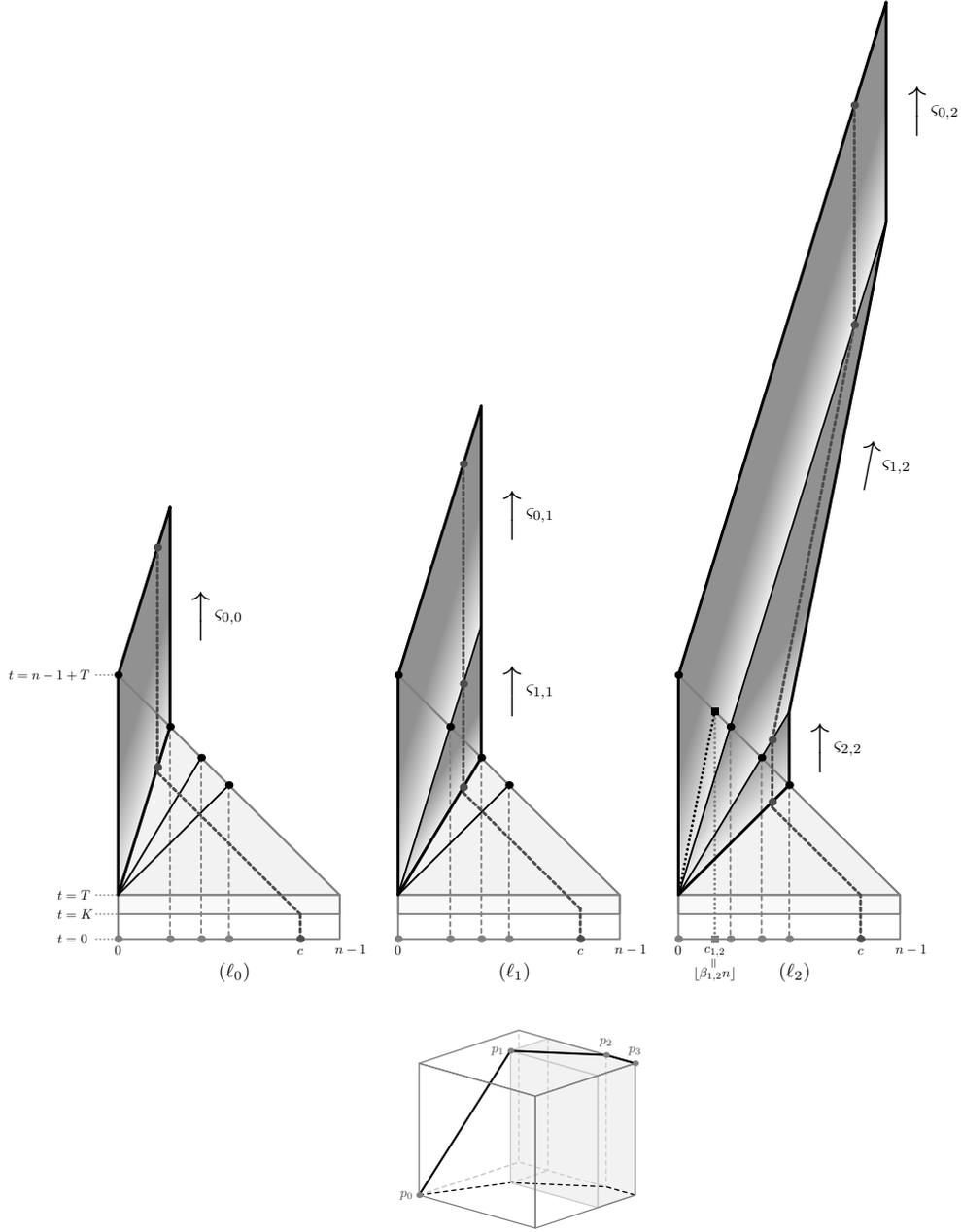

	\centering
	\begin{tabular}{ccc}
		\scalebox{.75}{\includegraphics{Pix/pix.7}} & \scalebox{.75}{\includegraphics{Pix/pix.6}} & \scalebox{.75}{\includegraphics{Pix/pix.5}}\\
		\\
		 & \scalebox{.75}{\includegraphics{Pix/pix.8}} &\\
	\end{tabular}
	\caption{Different compressed copies of the input shifted over $S$, with the trajectory of the letter initially contained by some cell~$c$ displayed. Layer~$\ell_j$ corresponds to head~$j < k = 3$. In this example, we have $(\alpha_{1, 1}, \alpha_{1, 2}, \alpha_{2, 2}) = (\frac 5 8, \frac 1 4, \frac 3 4)$. Hence, $(\alpha_0, \alpha_1, \alpha_2, \alpha_3) = (0, \frac{15}{64}, \frac 3 8, \frac 12)$ and $\beta_{1, 2} = \frac{21}{128}$.
	}
	\label{shift}
\end{figure}

Consider some head~$j\in\inter 0{k - 1}$ and an integer $i\in\inter 1j$. On layer~$\ell_j$, which  corresponds to this head, we want to shift the compressed input at some constant speed $\varsigma_{i, j}\in ]{-1}, \varsigma_i]$ between $D_{i + 1}$ and $D_i$, so that the correct letters pass over $S_i$. One can notice $\varsigma_{j, j} = 0$ by the definition of $\alpha_{j + 1}$ and $\alpha_j$. But this not necessarily the case when $i < j$. Indeed, $\varsigma_{i, j}$ should be defined as equal to $\cramped{\frac{\beta_{i, j}}{1 - \beta_{i, j}}}$, with $\beta_{i, j} = \alpha_i - \alpha_{i, j}\cramped{\frac{\alpha_{i + 1} - \alpha_i}{\alpha_{i + 1, j} - \alpha_{i, j}}}$ if $\alpha_{i + 1, j} - \alpha_{i, j} > 0$ and $\beta_{i, j} = \alpha_i$ otherwise. This way, $\varsigma_{i, j}$ is the speed of the signal we would use to mark cell~$c_{i, j} = \lfloor\beta_{i, j}n\rfloor$ (cf. fig.~\ref{shift}).

\subsection{Backtracking}\label{ss:s:rev}

Now that we have ensured the correct letters are seen in reverse order for each head on each segment $S_i$, how do we get site~$s_0$ to know the result of the execution of $\DFA$ over $w$? All we need to know is whether the final state of $\DFA$ is accepting, that is, belongs to $\Qa$.

Let $p$ be a point of $P$ such that $p\neq p_0$. One can observe that if we know $q$, the state $\DFA$ is in when its multi-head is on $p$, as well as the letter~$l_j\in\Sigma$ each head~$j$ reads when the multi-head lies on the predecessor $p'$ of $p$, then we can compute the possible states of $\DFA$ at point~$p'$. That is, the subset $Q'$ of $Q$ such that for all $q'\in Q$, $\delta(q', l_0, \dots, l_{k - 1}) = (q, p - p')\Leftrightarrow q'\in Q'$. Likewise, if we know $\DFA$ is in a state of $Q''\subseteq Q$ at point~$p$, we can determine the subset $Q'$ such that for all $q'\in Q$, $\delta(q', l_0, \dots, l_{k - 1})\in Q''\times\{p - p'\}\Leftrightarrow q'\in Q'$. We will refer to this process as \emph{reading $\delta$ backward}.

Let then $s = (c + 1, t - 1)$ be a site of $S$. As the letters it sees come from compressed versions of $w$, it can represent a (finite) range of points of $P$ instead of only one, depending on the part $S_i$ it belongs to. Now suppose it contains some subset of $Q$ for each of the successive points of $P$ it represents. Suppose also these subsets are consistent with one another (regarded as the possible states $\DFA$ is in at each of these points). Then successor site~$s' = (c, t)$ can read $\delta$ backward a finite (but sufficient) number of times to get the possible subsets of its own points.

Site~$s_k$ represents the last points of $P$, amongst which the very last point $p_k$. So, we initiate our `reverse' computation by setting the state of $s_k$ (on some layer~$\ell$ on which this computation is to be held) to contain subset~$\Qa$ for point~$p_k$ and consistent ones for the predecessors it represents. By induction, every element of $S$ will contain subsets that are consistent with $\Qa$ on layer~$\ell$. In particular, $s_0$ will have the corresponding subset $Q_0$ for $p_0$, so that it just has to check whether $q_0\in Q_0$ to know if $w$ is accepted by $\DFA$.\qed

\subsection{Adjustments}\label{ss:s:adjust}

In the preceding construction, we have put some details or particular cases aside. First, we have to mention that the whole process obviously works only for input words of size greater than some value depending on $K$ (for all $P_i$ to be assimilated to straight lines as in fig.~\ref{band}). Nevertheless, that leaves us a finite number of words that are treated as special cases, so that the result is not affected.

\subsubsection*{Possibilities}

As each $P_i$ is not a real straight line, the next part $P_{i + 1}$ of the path depends on which point of the period of $P_i$ the multi-head is at (that is, which state it is in over word~$a^n$) when head~$i$ reaches the end-marker. In particular, there can be at most $|Q|$ possible values $\alpha_i$, depending on $n$. Anyway, that makes a finite number of possible $(k - 1)$-tuples $(\alpha_1, \dots, \alpha_{k - 1})$, and we can thus process all of them in parallel.

Remains to elect the right tuple at site~$s_0$ or before. It can be done by looking at the remainder of the Euclidean division of $n - |Q|$ by some finite value $f(|Q|)$. That can be easily checked, for instance, on line $D_k$ with a finite counter. The choice will be known at site~$s_k$ and spread toward $s_0$.

\subsubsection*{Aperiodic parts}

It may seem we know at any site along any $S_i$, thanks to what precedes, which points of the period of $P_i$ we are simulating and so, which available letters the cell has to use. This is in fact not true yet: when reaching site~$s_i$, we have to take the aperiodic part of $P_i$ into account, and therefore we must be able to modify the last $|Q|$ moves (that is, to adjust the choice of letters) we have simulated backward. That can be done by adding to the sites of $S$ a finite memory of the letters seen.

\subsubsection*{Immobile heads}

Suppose that, contrary to the hypothesis made in subsection~\ref{ss:s:key}, there exists some $j > 0$ such that $\alpha_{j, j} = 0$. That means that head~$j$ remains motionless until $p_j$ and then covers the totality of the input during $P_j$. The trouble is that it implies for all $i\leq j$, $\alpha_i = 0$. Therefore, $s_j = s_{j - 1} = \dots = s_0$, so that a linear number of moves would have to be simulated on a single site.

A simple trick allows us to overcome this problem: for all $j\in\inter 1{k - 1}$, we set $\alpha'_{j, j} = \alpha_{j, j}$ if $\alpha_{j, j} > 0$ and $\alpha'_{j, j} = \frac 12$ otherwise\footnote{Notice we could have chosen any rational value strictly between $0$ and $1$ instead of $\frac 12$.}, and set $\alpha'_{0, 0} = \alpha_{0, 0} = 0$. Then, in our construction, we replace any $\alpha_{j, j}$ by $\alpha'_{j, j}$.

Finally, for each $j > 0$ verifying $\alpha_{j, j} = 0$, we still have to adjust shift speed~$\varsigma_{j, j}$, which is equal to $0$. All we have to do is to replace it by $\varsigma'_{j, j} = \cramped{\frac{\alpha_j}{1 - \alpha_j}}$ (only for this~$j$), which makes the totality of the copy of $w$ on layer~$\ell_j$ shift over $S_j$. As regards indices $i < j$, we do not need to redefine the corresponding speed $\varsigma_{i, j}$, since head~$j$ makes no more moves.

\section*{Conclusion}\label{s:concl}

We have described a construction that simulates oblivious multi-head one-way finite automata on real-time cellular automata. This is better (if linear and real times are not equivalent) than what would achieve the naïve (though nontrivial) simulation of general multi-head finite automata, which would result in a linear-time CA.

In any case, this result fully exploits the obliviousness of the sequential computation. Now, it is another challenge to get a similar parallel algorithm without the constraint of data-independence.

\section*{Acknowledgement}\label{s:ackn}

I would like to thank G.~Richard and V.~Terrier for introducing me to the matter of DIDFA (which resulted in a common article about a speed-up of two-way DIDFA by CA). I would also like to thank J.~Ferté for useful brainstorming sessions before the blackboard and V.~Poupet for his help.

\bibliographystyle{plain}
\bibliography{Acc1}

\end{document}